\documentclass[onecolumn,showpacs,amsmath,amssymb,pra]{revtex4}
\usepackage{graphicx}

\usepackage{amsmath,amssymb}

\begin{document}

\title{Schmidt modes and entanglement of biphoton polarization qutrits}

\author{Mikhail Fedorov}
\email{fedorov@ran.gpi.ru}
\author{Nikolai Miklin}
 \affiliation{A.M. Prokhorov General Physics Institute of the Russian Academy of Sciences,\\ 38 Vavilova st., Moscow 119991 Russia\\
Moscow Institute of Physics and Technology}

\date{\today}

\begin{abstract}
Polarization features and entanglement of biphoton polarization qutrits are briefly outlined. Schmidt modes of qutrits are found analytically and in a general form by the method different from the standard one and based on the original approach of Erhard Schmidt (1906).
\end{abstract}

\pacs{03.67.Bg, 03.67.Mn, 42.65.Lm}

\maketitle

\def\thesection{\arabic{section}}

\section{Schmidt decompositions}

Schmidt modes and Schmidt decompositions of pure bipartite states are important elements of the modern science of quantum information. On the other hand, complete mathematical formulation of these concepts was given more than hundred years ago by Erhard Schmidt in 1906 \cite{Schm}, before appearance of quantum mechanics and the concept of entanglement \cite{Schr}. Nowadays the Schmidt decomposition is understood as a feature of bipartite wave functions $\Psi(x_1,x_2)$ consisting in a possibility of presenting them by a sum of products of single-particle functions called Schmidt-modes
\begin{equation}
 \label{decomp}
 \Psi(x_1,x_2) = \sum_n\sqrt{\lambda_n}\,\psi_n(x_1)\chi_n(x_2),
\end{equation}
where $n$ is an integer, $n=1,2,...$, $x_1$ and $x_2$ are arbitrary variables of two particles (coordinates, or momenta, or even discrete variables such as polarizations of photons or spin projections of electrons, etc.).  Both the bipartite wave function $\Psi(x_1,x_2)$ and Schmidt modes are assumed to be unit-normalized. In both sets of Schmidt modes $\{\psi_n\}$ and $\{\chi_n\}$  all functions $\psi_n$ and $\psi_{n^\prime}$ with differing $n$ and $n^\prime$, as well as $\chi_n$ and $\chi_{n^\prime}$,  are orthogonal to each other. Owing to this, the real and positive parameters $\sqrt{\lambda_n}$ of the decomposition (\ref{decomp}) obey the normalization condition $\sum_n\lambda_n=1$. Both sets of functions $\{\psi_n\}$ and $\{\chi_n\}$ are complete and together they provide the only two-party basis, unique for any given pure bipartite state. Indeed, in principle, any function of two variables can be expanded in a double series of products of any single-particle functions if the sets of these functions are complete. Such expansions will contain two summations, over $n_1$ and $n_2$, corresponding to variables  $x_1$ and $x_2$. Only if the single-particle functions in  these expansions are the Schmidt modes, the double sums are reduced to single ones, as shown in Eq. (\ref{decomp}), and this makes the Schmidt-mode bases really unique.

The remarkable feature of Schmidt decompositions is related to the zero or 100$\%$ conditional probabilities of photon distributions in Schmidt modes. If we have a pair of particles, and we find, e.g., from experiment that one of them is in the Schmidt mode $\psi_n$, this guarantees automatically that the second particle of the same pair is in the adjoint mode with the same number $n$, $\chi_n$, and not in any other Schmidt mode. Entanglement of bipartite states is related to uncertainty of localization of pairs of particles in  different pairs of adjoint Schmidt modes. Probabilities of finding pairs of particles in the modes $\{\psi_n,\chi_n\}$  are given by the squared parameters of the Schmidt decomposition $\lambda_n$. The degree of entanglement is determined by the amount of efficiently populated Schmidt modes or, in other words, by the amount of sufficiently large decomposition parameters $\sqrt{\lambda_n}$. A natural characteristics of the degree of entanglement is provided by the Schmidt entanglement parameter (or Schmidt number) $K$ \cite{GRE,Knight} defined as
\begin{equation}
 \label{K}
 K=\left(\sum_n\lambda_n^2\right)^{-1}\geq 1.
\end{equation}A state is disentangled if only one of all parameters $\lambda_n$ differs from zero and equals unit, $\lambda_n=\delta_{n,1}$, when $K=1$ and the bipartite wave function factorizes: $\Psi(x_1,x_2)=\psi_1(x_1)\times\chi_1(x_2)$. In all other cases all $\lambda_n$ are smaller than unit and at least two pairs of Schmidt modes are populated. Then $K>1$, the wave function $\Psi(x_1,x_2)$ does not factorize, and the state is entangled.

The Schmidt decomposition (\ref{decomp}) takes somewhat simpler form if particles under consideration are indistinguishable, especially if they are bosons, like photons. The wave functions of pure biphoton states are obliged to be symmetric  with respect to the variable transpositions, $\Psi(x_1,x_2)=\Psi(x_2,x_1)$. Owing to this, the adjoint Schmidt modes of biphoton states $\psi_n$ and $\chi_n$ must be identical to each other to give
\begin{equation} \label{decomp-bf}
\Psi^{biph}(x_1,x_2) = \sum_n\sqrt{\lambda_n}\,\psi_n(x_1)\psi_n(x_2).
\end{equation}
As both photons in any pair of adjoint Schmidt modes are, in fact, in the same $n$-th mode, the Schmidt decomposition for the biphoton wave function (\ref{decomp-bf}) can be transformed into the Schmidt decomposition of  the biphoton state vector \cite{Ch-F}
\begin{equation}
 \label{decomp-st-vect}
 \left|\Psi^{biph}\right> = \sum_n\sqrt{\frac{\lambda_n}{2}}\,a_n^{\dag^2}|0>,
\end{equation}
where $a_n^\dag$ are the creation operators of photons in the Schmidt modes and $\Psi^{biph}(x_1,x_2)\equiv\langle x_1,x_2|\Psi^{biph}\rangle$.

\section{Definition and derivation of Schmidt modes}

The usual way of finding Schmidt modes consists in using the bipartite wave function $\Psi(x_1,x_2)$ of two variables $x_1$ and $x_2$ to construct the associated density matrix $\rho(x_1,x_2,x_1^\prime,x_2^\prime) = \Psi(x_1,x_2)\Psi^*(x_1^\prime,x_2^\prime)$. Then the two reduced density matrices are defined as traces of $\rho$ over one of two variables, $\rho_r^{(1)}(x_1,x_1^\prime)=\int dx_2\rho(x_1,x_2,x_1^\prime,x_2)$ and $\rho_r^{(2)}(x_2,x_2^\prime)=\int dx_1\rho(x_1,x_2,x_1,x_2^\prime)$. After this, the Schmidt modes $\psi_n(x_1)$  and $\chi_n(x_2)$ are defined as eigenfunctions of two integral equations kernels of which are given by the reduced density matrices
\begin{eqnarray}
 \label{int-eq-rho}
  \int dx_1^\prime \rho_r^{(1)}(x_1,x_1^\prime) \psi_n(x_1^\prime)=\lambda_n\psi_n(x_1),\quad
 \int dx_2^\prime \rho_r^{(2)}(x_2,x_2^\prime) \chi_n(x_2^\prime)=\lambda_n\chi_n(x_2).
\end{eqnarray}
By definition, the Schmidt modes diagonalize the reduced density matrices, and from here one derives the Schmidt decomposition (\ref{decomp}) of the wave function $\Psi(x_1,x_2)$ \cite{Peres}.

But the logic of E.Schmidt is different, and the consequence of actions is opposite to that described above. Adapted to the terminology of quantum mechanics, the first step to the Schmidt's derivation consists in finding Schmidt modes from the integral equations. the kernels of which are given by the wave function $\Psi(x_1,x_2)$
\begin{equation}
 \label{int-eqs}
 \int dx_2\Psi(x_1,x_2)\chi_n^*(x_2) = \sqrt{\lambda_n}\,\psi_n(x_1),\quad
 \int dx_1\psi_n^*(x_1)\Psi(x_1,x_2) = \sqrt{\lambda_n}\,\chi_n(x_2).
\end{equation}
In principle these equations follow from the Schmidt decomposition (\ref{decomp}), if the latter is assumed to be proved. This decomposition is used by E. Schmidt for proving that the Schmidt modes defined by Eqs. (\ref{int-eqs}) diagonalize the reduced density matrices. In the case ofbiphoton states, owing to the symmetry requirements, two equations (\ref{int-eqs}) are reduced to a single one
\begin{equation}
 \label{int-eq-symm}
 \int dx_1\psi_n^*(x_1)\Psi(x_1,x_2)=\sqrt{\lambda_n}\,\psi_n(x_2).
\end{equation}

Thus, the Schmidt modes can be found either from Eqs. (\ref{int-eq-rho}) or from Eqs. (\ref{int-eqs}), (\ref{int-eq-symm}). Though two methods are almost equivalent, the method based on the use of the bipartite wave function as the kernel of equations has at least one advantage. As Eqs. (\ref{int-eqs}), (\ref{int-eq-symm}) contain both the Schmidt modes themselves and their complex conjugate functions, these equations permit obtaining Schmidt modes with correctly determined phases, whereas solutions of Eqs. (\ref{int-eq-rho}) leave the phases of Schmidt modes undefined. For a single Schmidt mode taken as itself its phase can be not important. But as the Schmidt decomposition (\ref{decomp}) contains the superposition of products of Schmidt modes, their phases are crucially important. For this reason, the method based on the use of the bipartite wave function as the kernel of equations is preferable for correct and complete derivation of Schmidt modes. Below we show how this method works in the case of biphoton polarization qutrits.

\section{Biphoton polarization qutrits and their properties}

As known, biphoton polarization qutrits are states of two collinearly propagating photons, both in the same spatial-spectral mode but with either different or coinciding polarizations. Polarization is the only degree of freedom in which such states can be entangled. In a general form, the state vector of biphoton polarization qutrits is given by
\begin{equation}
 \label{QTR}
 |\Psi\rangle= C_1|2_H\rangle+C_2|1_H,1_V\rangle+C_3|2_V\rangle
 \equiv
\left(C_1\frac{a_H^{\dag^{\,2}}}{\sqrt{2}}+C_2\,a_H^\dag  a_V^\dag+C_3\frac{a_V^{\dag^{\,2}}}{\sqrt{2}}\right)|0\rangle,
\end{equation}
where $a_H^\dag$ and $a_V^\dag$ are the creation operators of  a photon in the modes with horizontal ($H$) and vertical ($V$) polarizations. $C_{1,2,3}$ are arbitrary complex constants restricted by the normalization condition $\sum_i|C_i|^2=1$.  Experimentally such general-form biphoton polarization qutrits can be produced in processes of Spontaneous Parametric Down-Conversion (SPDC) either with three nonlinear crystals installed closely one after another or with the help of manipulations with polarizations after formation in a single crystal.

Three terms in both expressions of Eq. (\ref{QTR}) are basis state vectors. Similarly to (\ref{QTR}), the qutrit's wave function is given by a superposition of three basis wave functions
\begin{equation}
 \Psi(\sigma_1,\sigma_2)=C_1\psi_{HH}(\sigma_1,\sigma_2)+C_2\psi_{HV}(\sigma_1,\sigma_2)+C_3\psi_{VV}(\sigma_1,\sigma_2)
 \label{QTR-wf}
\end{equation}
with explicit expressions for basis wave functions given by
\begin{eqnarray}
 \nonumber
 \psi_{HH}(\sigma_1,\sigma_2) =\delta_{\sigma_1,H}\delta_{\sigma_2,H} \equiv\left({1\atop 0}\right)_1\otimes\left({1\atop 0}\right)_2,\quad\psi_{VV}(\sigma_1,\sigma_2) =\delta_{\sigma_1,V}\delta_{\sigma_2,V}\equiv\left({0\atop 1}\right)_1\otimes\left({0\atop 1}\right)_2,\\
 \label{basis wf}
 \psi_{HV}(\sigma_1,\sigma_2) = \frac{1}{\sqrt{2}}\Big(\delta_{\sigma_1,H}\delta_{\sigma_2,V}+ \delta_{\sigma_1,V}\delta_{\sigma_2,H}\Big)\equiv\frac{1}{\sqrt{2}}\left[\left({1\atop 0}\right)_1\otimes\left({0\atop 1}\right)_2+\left({0\atop 1}\right)_1\otimes\left({1\atop 0}\right)_2\right].
\end{eqnarray}
Here and below the upper and lower lines in columns correspond to the horizontal and vertical polarizations, subscripts 1 and 2 correspond to variables $\sigma_1$ and $\sigma_2$ of two indistinguishable photons, and the obligatory symmetry of biphoton wave function is taken into account explicitly in $\psi_{HV}$. The functional and matrix forms of writing wave functions are absolutely equivalent as, for example, $\left(1\atop 0\right)_1\equiv\delta_{\sigma_1,H}$, $\left(0\atop 1\right)_2\equiv\delta_{\sigma_2,V}$, etc.

The first two basis wave functions in Eq. (\ref{basis wf}) ($\psi_{HH}$ and $\psi_{VV}$), are factored
whereas the third one ($\psi_{HV}$) is unfactorable. This means that the states $|2_H\rangle$ and
$|2_V\rangle$ are disentangled and the state $|1_H,1_V\rangle$ is entangled. For the qutrit state of  a general form
(\ref{QTR}) its concurence was found  in the work \cite{NJP} with the help of the Wootters procedure \cite{Wootters} to be given by
\begin{equation}
 \label{C}
 C = \left|2C_1C_3-C_2^2\right|.
\end{equation}
As known \cite{Rungta}, for two-qubit states there is the following simple relation between the concurrence $C$ and Schmidt entanglement parameter $K$: $C=\sqrt{2(1-K^{-1})}$, which gives
\begin{equation}
 \label{K}
 K = \frac{2}{2-\left|2C_1C_3-C_2^2\right|^2}.
\end{equation}
In dependence on values of the constants $C_{1,2,3}$, parameters $C$ and $K$ vary synchronously in the limits $0\leq C\leq 1$ and $1\leq K\leq 2$. The lower limits correspond to disentangled and the upper - to maximally entangled states of biphoton polarization qutrits. In particular, as was found from structures of biphoton basis wave function, examples of disentangled states are $|2_H\rangle$ ($C_2=C_3=0$) and $|2_V\rangle$ ($C_2=C_1=0$). In both cases $C=0$ and $K=1$. An example of the maximally entangled state is the state $|1_H,1_V\rangle$  ($C_1=C_3=0$) for which $C=1$ and $K=2$. In a more general form, arbitrary disentangled states have the form $A^{\dag^2}|0\rangle/\sqrt{2}$ with arbitrary one-photon creation operators $A^\dag$, and all maximally entangled states have the form $A^\dag B^\dag|0\rangle$ with $A^\dag$ and $B^\dag$ being creation operators of photons in arbitrary orthogonal modes (with zero commutator of operators $A$ and $B^\dag$, $[A,B^\dag]=0$) \cite{Ch-F}. All other qutrits are partially entangled and are characterized by intermediate values of parameters $C$ and $K$, $C\neq 0\,{\rm or}\, 1$, $K\neq 1\,{\rm or}\,2$.

\section{Derivation of Schmidt modes of biphoton polarization qutrits}

As explained in section 2, Schmidt modes of any biphoton state can be completely determined from the integral equation of the form (\ref{int-eq-symm}) with the kernel given by the biphoton wave function. In the case of polarization qutrits the variables of the wave function are discrete, and the integral equation for finding Schmidt modes takes the form of a ``sum equation$"$
\begin{equation}
 \label{int-discr-eq}
 \sum_{\sigma_1}\psi^*(\sigma_1)\Psi(\sigma_1,\sigma_2)=\sqrt{\lambda}\,\psi(\sigma_2).
\end{equation}
It's most convenient to rewrite this equation in the matrix form by using matrix expressions (\ref{basis wf}) for basis wave functions and their sum $(\Psi)$ with coefficients $C_{1,2,3}$. The Schmidt modes in the matrix form can be written as  $\psi=\left(\alpha\atop\beta\right)$ with the unknown complex constants $\alpha=|\alpha|e^{i\varphi_\alpha}$ and $\beta=|\beta|e^{i\varphi_\beta}$ obeying the normalization condition $|\alpha|^2+|\beta|^2=1$. Then Eq. (\ref{int-discr-eq})
 takes the form
\begin{equation}
 \label{QTR eq for Schm modes}
 \left(\alpha^*\;\beta^*\right)_1\cdot (\Psi)=\sqrt{\lambda}\,\left({\alpha\atop \beta}\right)_2,
\end{equation}
which yields the following two equations for the constants $\alpha$ and $\beta$
\begin{equation}
 \label{eq for a and b}
 \left\{
 \begin{matrix}
 C_1\,\alpha^*+\displaystyle\frac{C_2}{\sqrt{2}}\,\beta^*=\sqrt{\lambda}\,\alpha,\\
\displaystyle\frac{C_2}{\sqrt{2}}\,\alpha^*+C_3\,\beta^*=\sqrt{\lambda}\,\beta.
 \end{matrix}
 \right.
\end{equation}
These equations and their solution are very simple. By assuming that $C_2\neq 0$, we can use the first of Eqs. (\ref{eq for a and b}) to express $\beta$ and $\beta^*$ in terms of $\alpha$ and $\alpha^*$:
\begin{equation}
 \label{b via a}
 \beta^*=\frac{\sqrt{2}}{C_2}\left[\sqrt{\lambda}\,\alpha-C_1\,\alpha^*\right],\,\beta=\frac{\sqrt{2}}{C^*_2}\left[\sqrt{\lambda}\,\alpha^*-C_1^*\,\alpha\right].
\end{equation}
By substituting these expressions into the second of Eqs. (\ref{eq for a and b}) we find the ratio of $\alpha$ to $\alpha^*$
\begin{equation}
 \label{varphi-a}
 \frac{\alpha}{\alpha^*}\equiv e^{2i\varphi_\alpha}=\frac{C_2^*(2C_1C_3-C_2^2)+2\lambda C_2}{2\sqrt{\lambda}(C_2^*C_3+C_2C_1^*)}.
\end{equation}
Similarly, we can use the the first of Eqs. (\ref{eq for a and b}) to express $\alpha^*$ via $\beta$ and $\beta^*$. Substituted to the second of Eqs. (\ref{eq for a and b}), this gives the ratio of $\beta$ to $\beta^*$, differing from $\alpha/\alpha^*$ ({\ref{varphi-a}) only by the transposition $C_1\rightleftharpoons C_3$
\begin{equation}
 \label{varphi-b}
 \frac{\beta}{\beta^*}=e^{2i\varphi_\beta}=\frac{C_2^*(2C_1C_3-C_2^2)+2\lambda C_2}{2\sqrt{\lambda}(C_2^*C_1+C_2C_3^*)}.
\end{equation}
As absolute values of the ratios $\alpha/\alpha^*$ and $\beta/\beta^*$ are equal unity, e.g., from Eq. (\ref{varphi-a}) we get
\begin{equation}
 \label{Eq. for lambda}
 \left|\frac{\alpha}{\alpha^*}\right|=\left|\frac{C_2^*(2C_1C_3-C_2^2)+2\lambda C_2}{2\sqrt{\lambda}(C_2^*C_3+C_2C_1^*)}\right|=1.
\end{equation}
This is the equation for $\lambda$ from which we find with a simple algebra that it has only two nonzero solutions given by
\begin{equation}
 \label{lambda pm}
 \lambda_\pm=\frac{1}{2}\left(1\pm\sqrt{1-C^2}\right),
\end{equation}
where $C$ is the qutrit's concurrence of Eq. (\ref{C}). The parameters $\lambda_\pm$ (\ref{lambda pm}) coincide with the eigenvalues of the qutrit reduced density matrix found earlier \cite{NJP} by absolutely different methods. These eigenvalues correspond to two Schmidt modes of qutrits
\begin{equation}
 \label{two-Schm-modes}
 \psi_\pm=\left(\alpha_\pm\atop{\beta_\pm}\right)
 =\left(|\alpha_\pm|e^{i\varphi_{\alpha\,\pm}}\atop{|\beta_\pm|e^{i\varphi_{\beta\,\pm}}}\right).
\end{equation}
With $\lambda_\pm$ (\ref{lambda pm}) substituted into Eqs. (\ref{varphi-a}) and (\ref{varphi-b}), one can use these equations for finding  directly phases $\varphi_{\alpha\,\pm}$ and $\varphi_{\beta\,\pm}$ as one half of phases of expressions on the right-hand sides of these equation.

For finding absolute values of the constants $\alpha_\pm$ and $\beta_\pm$, we can use, e.g., the first of Eqs. (\ref{b via a}), which yields
\begin{equation}
 \frac{|\beta_\pm|}{|\alpha_\pm|}=\frac{\sqrt{2}}{|C_2|}\,\Big|\sqrt{\lambda_\pm}e^{2i\varphi_{\alpha\,\pm}}-C_1\Big|
 \label{b divide a}
 =\frac{\left||C_1|^2-|C_3|^2\mp\sqrt{1-C^2}\,\right|}{\sqrt{2}\,\left|C_2^*C_3+C_2C_1^*\right|}.
\end{equation}
Eq. (\ref{b divide a}), as well as the final expressions for the Schmidt-mode parameters $\alpha_\pm$ and $\beta_\pm$, can be significantly simplified with the help of the following `magic' relation between the constants $C_{1,2,3}$ and concurrence $C$ (\ref{C})
\begin{equation}
 \label{magic}
2\left|C_2^*C_3+C_2C_1^*\right|^2=-\left(|C_1|^2-|C_3|^2\right)^2+1-C^2\equiv (1-C^2)(1-x^2),
\end{equation}
where $x$ is one of the main parameters determining the structure of the qutrit's Schmidt modes
\begin{equation}
 \label{x}
  x=\frac{|C_1|^2-|C_3|^2}{\sqrt{1-C^2}}.
\end{equation}
As follows from Eq. (\ref{magic}), at any values of the constants $C_{1,2,3}$, the absolute value of the parameter $x$ does not exceed unit, $0\leq |x|\leq 1$. In terms of $x$, Eq. (\ref{b divide a}) takes a much simpler form
\begin{equation}
 \frac{|\beta_\pm|}{|\alpha_\pm|}=\sqrt{\frac{1\mp x}{1\pm x}}.
 \label{b div a via x}
\end{equation}
With the normalization condition taken into account, absolute values of the constants $\alpha_\pm$ and $\beta_\pm$ are easily found to be given by
\begin{equation}
 \label{|a|,|b|}
 |\alpha_\pm|=\sqrt{\frac{1\pm x}{2}},\, |\beta_\pm|=\sqrt{\frac{1\mp x}{2}}.
\end{equation}

Note that Eqs. (\ref{varphi-a}) and (\ref{varphi-b}) can make impression that their solutions $\varphi_{\alpha\,\pm}$ and $\varphi_{\beta\,\pm}$ can be independently shifted by $\pi$ or $-\pi$, i.e., that signs of the phase factors $e^{i\varphi_{\alpha\,\pm}}$ and $e^{i\varphi_{\beta\,\pm}}$ remain absolutely uncertain. In fact this is true only for one of the phases, e.g., $\varphi_{\alpha\,\pm}$. At a given $\varphi_{\alpha\,\pm}$ the phase $\varphi_{\beta\,\pm}$ has to be strictly defined. This means that by using Eq. (\ref{varphi-a}) for finding $\varphi_{\alpha\,\pm}$, we have to use directly one of Eqs. (\ref{eq for a and b}) or (\ref{b via a}) for finding $\varphi_{\beta\,\pm}$, rather than Eq. (\ref{varphi-b}). Specifically, for finding the difference of phases $\varphi_{\beta\,\pm}-\varphi_{\alpha\,\pm}$, the second of Eqs. (\ref{b via a}) can be used to give
\begin{equation}
 \label{phib-phia}
 e^{i(\varphi_{\beta\,\pm} - \varphi_{\alpha\,\pm})} = \sqrt{2}\frac{|\alpha_\pm|}{|\beta_\pm|} \\
 \times  \left(\frac{\sqrt{\lambda_\pm}e^{2i\varphi_{\alpha\,\pm}}-C_1}{C_2}\right)^* \\
 = \frac{|\alpha_\pm|}{\sqrt{2}\,|\beta_\pm|}\frac{|C_3|^2-|C_1|^2\pm\sqrt{1-C^2}}{C_1C_2^*+C_2C_3^*} \\
 = \pm\, e^{i\varphi_0},
\end{equation}
where
\begin{equation}
 \label{phi-0}
 \varphi_0=arg\left(C_1^*C_2+C_2^*C_3\right).
\end{equation}
Thus, Eqs. (\ref{phib-phia}) and (\ref{phi-0}) yield
\begin{equation}
 \label{phib via phia}
 \varphi_{\beta\,+}=\varphi_{\alpha\,+}+\varphi_0\quad{\rm and}\quad\varphi_{\beta\,-}=\varphi_{\alpha\,-}+\varphi_0+\pi .
\end{equation}
Eqs. (\ref{|a|,|b|}) and (\ref{x}) together with Eqs. (\ref{varphi-a}) and (\ref{phi-0}), (\ref{phib via phia}) conclude the derivation of expressions for the Schmidt modes of biphoton polarization qutrits in a general form. The Schmidt decomposition for the general qutrit wave function is given by
\begin{equation}
 \label{Schm-decomp-wf-QTR}
 \Psi=\sum_\pm\sqrt{\lambda_\pm}\left(|\alpha_\pm|e^{\varphi_{\alpha\,\pm}}\atop{|\beta_\pm|e^{\varphi_{\beta\,\pm}}}\right)_1
 \otimes\left(|\alpha_\pm|e^{\varphi_{\alpha\,\pm}}\atop{|\beta_\pm|e^{\varphi_{\beta\,\pm}}}\right)_2.
\end{equation}
In accordance with the general features of biphoton states (\ref{decomp-st-vect}), the decomposition of the wave function(\ref{Schm-decomp-wf-QTR}) corresponds to the Schmidt decomposition of the general qutrit state vector
\begin{equation}
 \label{Schm-decomp-st-vect-QTR}
 |\Psi\rangle=\left(\sqrt{\frac{\lambda_+}{2}}\,a_+^{\dag^{\,2}}+\sqrt{\frac{\lambda_-}{2}}\,a_-^{\dag^{\,2}}\right)|0\rangle,
\end{equation}
where $a_+^\dag$ and $a_-^\dag$ are the photon creation operators for the Schmidt modes $+$ and $-$
\begin{equation}
 \label{Schm-mode-cre--operators}
 a_\pm^\dag=\alpha_\pm\,a_H^\dag+\beta_\pm\,a_V^\dag.
\end{equation}

\section{Examples}
\subsection{$C_2=0$}
In principle, this case is so simple that it needs almost no derivations. Indeed, in the case $C_2=0$ the expression (\ref{QTR}) for the qutrit's state vector can be rewritten as
\begin{equation}
 \label{QTR-C2=0}
 |\Psi\rangle=\left(|C_1|e^{i\varphi_1}\frac{a_H^{\dag^2}}{\sqrt{2}}
 +|C_3|e^{i\varphi_3}\frac{a_V^{\dag^2}}{\sqrt{2}}\right)|0\rangle,
\end{equation}
where $\varphi_{1,3}$ are phases of the constants $C_{1,3}$. Comparison with Eq. (\ref{Schm-decomp-st-vect-QTR}) gives immediately expressions for the eigenvalues of the ``integral" equation (\ref{QTR eq for Schm modes})
\begin{equation}
\label{C2=0 lambda }
 \lambda_+=|C_1|^2,\;\lambda_-=|C_3|^2\\
\end{equation}
and for the Schmidt-mode creation operators
\begin{equation}
 \label{C2=0 schm-mode-oper}
 a_+^\dag=e^{i\varphi_1/2}a_H^\dag,\;a_-^\dag=e^{i\varphi_3/2}a_V^\dag.
\end{equation}
The Schmidt modes themselves are given by
\begin{equation}
 \label{C2=0 schm-modes}
 \psi_+=\left({e^{i\varphi_1/2}\atop 0}\right),\;\psi_-=\left({0\atop e^{i\varphi_3/2}}\right).
\end{equation}
Though this derivation is very simple and sufficient, it can be interesting to see how the same results follow from the general expressions of the Schmidt modes in the case $C_2=0$. As in this case, in accordance with Eq.(\ref{lambda pm}), $C=2|C_1||C_3|$, and $C_1|^2+|C_3|^2=1$, the parameter $x$ of Eq. (\ref{x}) equals $1$ or $-1$ if, correspondingly, $|C_1|>|C_3|$ or $|C_1|<|C_3|$. These two cases are identical  to each other with substitutions $\lambda_+\rightleftharpoons \lambda_-$. For this reason, let us take $|C_1|>|C_3|$. Then Eqs. (\ref{|a|,|b|}) with $x=1$ give $|\alpha_+|=1,\,|\beta_+|=0$ and $|\alpha_-|=0,\,|\beta_-|=1$ in agreement with Eqs. (\ref{C2=0 schm-modes}). As for the phases $\varphi_{\alpha,\beta}$, both numerators and denominators of expressions in Eqs. (\ref{varphi-a}) and (\ref{varphi-b}) turn zero at $C_2=0$; to define their ratios, we have to take small but finite $C_2$ and to  approximate numerators by terms linear in $C_2$. After this we get the following expressions for the phase factors of the nonzero components of the Schmidt modes:
\begin{equation}
 \label{phases-C2=0}
 e^{2i\varphi_{\alpha\,+}}=\frac{C_1}{|C_1|}=e^{i\varphi_1},\,e^{2i\varphi_{\beta\,-}}=\frac{C_3}{|C_3|}=e^{i\varphi_3},
\end{equation}
i.e., $\varphi_{a\,+}=\varphi_1/2$ and $\varphi_{b\,-}=\varphi_3/2$, which also agrees with the results of Eqs. (\ref{C2=0 schm-modes}).

\subsection{$C_1=C_3=0,\,C_2=1$}
This is the case of the state $|1_H,1_V\rangle=a_H^\dag a_V^\dag |0\rangle$, most often raising discussions on whether this state is entangled or not. As shown above, this state is entangled because its wave function $\psi_{HV}$ is unfactorable [the last formula in Eqs. (\ref{basis wf})] and its concurrence $C$ (\ref{C}) equals unit. The Schmidt modes of the state $|1_H,1_V\rangle$ have been found \cite{NJP} to be given by $|45^\circ\rangle$ and $i|135^\circ\rangle$, where $45^\circ$ and $135^\circ$ are two orthogonal directions of polarization turned for $45^\circ$ with respect to the horizontal and vertical axes. It's rather interesting to see how these results follow from the general equations derived above [(\ref{varphi-a}), (\ref{varphi-b}) and (\ref{|a|,|b|}), (\ref{x})].

If both constants $C_1$ and $C_3$ are equal zero exactly, then $C=2\lambda_\pm=1$ (\ref{lambda pm}), and the fraction in Eq. (\ref{x}) is not defined: both its numerator and denominator turn zero. To define the value of the parameter $x$ (\ref{x}), we can assume that $C_3\equiv 0$, whereas $C_1\neq 0$ but small and, e.g.,  real. In the lowest order in $C_1$ we have $C^2=(1-|C_1|^2)^2\approx 1-2|C_1|^2$ and $\sqrt{1-C^2}\approx \sqrt{2}|C_1|$. Thus, the denominator of the fraction in Eq. (\ref{x}) is linear in $C_1$, whereas the numerator is quadratic. In the limit $C_1\rightarrow 0$ Eq. (\ref{x}) gives $x=0$. With $x=0$, Eqs. (\ref{|a|,|b|}) give $|\alpha|=|\beta|=\frac{1}{\sqrt{2}}$. The phases $\varphi_{\alpha\,\pm}$ can be found from Eq. (\ref{varphi-a}) with the help of the same limiting-transition procedure, which gives $2\lambda_\pm\approx 1\pm\sqrt{2}|C_1|$. Owing to this, we get $e^{2i\varphi_{\alpha\,\pm}}=\pm 1$ and, hence, the phases $\varphi_{\alpha\,\pm}$ can be taken equal to $\varphi_{\alpha\,+}=0$ and $\varphi_{a\,-}=-\frac{\pi}{2}$. The phases $\varphi_{\beta\,\pm}$ are determined by Eq. (\ref{phib via phia}) with $\varphi_0$ of Eq. (\ref{phi-0}) equal zero, which gives $\varphi_{\beta\,+}=\varphi_{\alpha\,+}=0$ and $\varphi_{\beta\,-}=\varphi_{\alpha\,-}+\pi=\pi/2$. Thus, the Schmidt modes of the state
$|1_H,1_V\rangle$ are\begin{equation}
 \label{HV-modes}
 \psi_+=\frac{1}{\sqrt{2}}\left({1\atop 1}\right),\;\psi_-=\frac{i}{\sqrt{2}}\left({-1\atop 1}\right),
\end{equation}
in complete agreement with the results of Ref. \cite{NJP}. Existence of two different Schmidt modes indicates directly that the wave function of the state $|1_H,1_V\rangle$ is unfactorable and the state is entangled.

\subsection{$C_3=0$}
In this case the state vector of the qutrit can be taken in the form
\begin{equation}
 \label{QTR-theta}
 |\Psi\rangle=\frac{a_H^\dag\,\big(\cos\theta\,a_H^\dag+e^{i\phi}\sin\theta\,a_V^\dag\big)}{\sqrt{1+\cos^2\theta}}|0\rangle
\end{equation}
with only two arbitrary and independent parameters $\theta$ and $\phi$. By comparing Eq. (\ref{QTR-theta}) with (\ref{QTR}), we see that in the case under consideration there are only two nonzero qutrit's parameter $C_1$ and $C_2$ equal to
\begin{equation}
 \label{C1-C2}
 C_1=\frac{\sqrt{2}\cos\theta}{\sqrt{1+\cos^2\theta}},\;C_2=\frac{e^{i\phi}\sin\theta}{\sqrt{1+\cos^2\theta}}.
\end{equation}
For this configuration Eqs. (\ref{C}), (\ref{lambda pm}) and (\ref{x}) give the following expressions for the concurrence $C$ and parameters $\lambda_\pm$  and $x$
\begin{equation}
 \label{C,lambda,x}
 C=\frac{\sin^2\theta}{1+\cos^2\theta},\;\lambda_\pm=\frac{(1\pm\cos\theta)^2}{2(1+\cos^2\theta)},\;x=\cos\theta.
\end{equation}
The absolute values of the Schmidt-mode components (\ref{|a|,|b|}) are easily found to be given by
\begin{equation}
 \label{|a|. |b| C3=0}
 |\alpha_+|=|\beta_-|=\cos(\theta/2),\,|\alpha_-|=|\beta_+|=\sin(\theta/2).
\end{equation}
The phases of $\alpha_\pm$ are determined by Eq. (\ref{varphi-a}), which after all substitutions, takes a very simple form $e^{2i\varphi_{\alpha\,\pm}}=\pm 1$. This gives
\begin{equation}
 \label{phi-a-c3=0}
 \varphi_{\alpha\,+}=0,\;\varphi_{\alpha\,-}=-\frac{\pi}{2}.
\end{equation}
The phases $\varphi_{\beta\,\pm}$ are determined by Eq. (\ref{phib via phia}) with Eq. (\ref{phi-0}) giving $\varphi_0=\phi$, which yields
\begin{equation}
 \label{phi-b-c3=0}
 \varphi_{\beta\,+}=\phi,\;\varphi_{\beta\,-}=\phi+\frac{\pi}{2}.
\end{equation}
Thus, the Schmidt modes of the qutrit (\ref{QTR-theta}) are given by
\begin{equation}
 \label{modes-C3=0}
 \psi_+=\left({\cos(\theta/2)\atop e^{i\phi}\sin(\theta/2)}\right),\;\psi_-=i\,\left({-\sin(\theta/2)\atop e^{i\phi}\cos(\theta/2)}\right).
\end{equation}
In the case $\phi=0$ these expressions coincide with those obtained in Ref. \cite{Ch-F} by the method absolutely different from the present one. We think that in this way these two methods confirm validity of both of them and support each other.

\section{Simplification of and manipulations with the Schmidt decomposition of biphoton qutrits}

A general Schmidt decomposition of the qutrit state vector (\ref{Schm-decomp-st-vect-QTR}) can be further simplified by means of transformations easily realizable experimentally. As known, with the help of properly installed half- and quarter-wavelength plates any photon polarization can be transformed, e.g., to the horizontal one. If this procedure is applied to the Schmidt mode $\psi_+$, the second Schmidt mode, $\psi_-$, transforms simultaneously into the state with vertical polarization, as the the described transformation conserves orthogonality of Schmidt modes. As the result the  qutrit's state vector   (\ref{Schm-decomp-st-vect-QTR}) takes the form
\begin{equation}
 \label{Schm-decomp-simplest}
 |\Psi_{simplest}\rangle=\left(\sqrt{\frac{\lambda_+}{2}}\,a_H^{\dag^{\,2}}+e^{2i\phi}\sqrt{\frac{\lambda_-}{2}}
 \,a_V^{\dag^{\,2}}\right)|0\rangle.
\end{equation}
The phase factor $e^{2i\phi}$ is not controlled by the described transformation and arises from phases of Schmidt modes. In fact, the phase $\phi$ is one of two remaining constants characterizing the qutrit's states in the form (\ref{Schm-decomp-simplest}). The second constant is any combination of $\lambda_+$ and $\lambda_-$, different from their sum equal unit. Found experimentally, $\lambda_+$ and $\lambda_-$ determine the degree of entanglement of any given qutrit, $C=2\sqrt{\lambda_+\lambda_-}$. A rather simple method of measuring both $\lambda_\pm$ and the phase $\phi$ was suggested in the work \cite{Ch-F}. The main idea is splitting the biphoton beam by the polarization beam splitter and making coincidence measurements in one of the channels. It appears useful using different orientations of the beam splitter: the usual one and, in the second series of measurements, turned for $90^\circ$ around the photon propagation axis. The same measurements with the beam splitter turned for $45^\circ$ were shown to be sufficient for measuring the phase $\phi$. Moreover, the phase $\phi$ is easily accessible for manipulations. The simplest way of changing $\phi$ is installing on the way of the biphoton beam  two birefringent plates turned for some angles in opposite directions. The arising states are of the same form as in Eq. (\ref{Schm-decomp-simplest}) but with $\phi$ substituted by $\phi+\Delta\phi$. Variations of the phase $\phi$ do not change the degree of either entanglement $C$ or polarization $P$, but they can be used for encoding and transmission of information. This encoding method was found to be crucially dependent on entanglement of biphoton qutrits: it can be most efficient in states with $C=1$, whereas in the case of disentangled states ($C=0$) the phases $\phi$ and $\phi+\Delta\phi$ are indistinguishable. The latter is clear already from the form (\ref{Schm-decomp-simplest}) of the Schmidt decomposition because the case $C=0$ corresponds to $\lambda_-=0$ when the state $|\Psi_{simplest}\rangle$ does not depend on $\phi$ at all. Thus, indeed,  the operational features of biphoton qutrits  are most pronounced when $C=1$, which occurs in states with two equally populated orthogonal Schmidt modes, such as $|1_H,1_V\rangle$, or $|1_{45^\circ},1_{135^\circ}\rangle$, etc.

\section{Conclusion}
In this work we have described a method of finding Schmidt modes of pure bipartite states different from the traditionally used one but agreeing completely with the logic of proofs in the original work by E. Schmidt \cite{Schm}. The method is based on defining Schmidt modes as eigenfunctions of integral equations with kernels given by the bipartite wave functions rather than reduced density matrices. This approach permits to find Schmidt modes completely with their phases expressed in terms of parameters characterizing the bipartite wave function. In contrast, in definitions with the reduced-density-matrix kernels phases of Schmidt modes remain undefined. The method of wave-function kernels is applied for finding Schmidt modes of biphoton polarization qutrits analytically and in a general form. The Schmidt decomposition of the wave function is transformed into the Schmidt decomposition of state vectors of biphoton qutrits. This form of the Schmidt decomposition is very convenient for further simplifications and manipulations. In particular, a simple experimental scheme is suggested in which the Schmidt decomposition of the qutrit state vector is reduced to the simplest form (\ref{Schm-decomp-simplest}) characterized only by two parameters: concurrence (the degree of entanglement) and the phase $\phi$ between two Schmidt modes. The schemes for direct experimental measurement of these parameters are suggested. Also, as shown, the phase $\phi$ can be easily changed experimentally to give rise to new states with the same degree of entanglement. A manifold of states with different values of the phase $\phi$ can be used for encoding and transmission of information, if only the degree of entanglement is high enough, which shows explicitly that biphoton polarization qutrits obey operational properties related to their entanglement.

\end{document}